# A Fast Image Encryption Scheme based on Chaotic Standard Map

Kwok-Wo Wong, Bernie Sin-Hung Kwok, and Wing-Shing Law

Department of Electronic Engineering,

City University of Hong Kong, 83 Tat Chee Avenue, Kowloon Tong, HONG KONG

E-mail: itkwwong@cityu.edu.hk (K.W. Wong)

**Abstract**

In recent years, a variety of effective chaos-based image encryption schemes have been proposed. The typical structure of these schemes has the permutation and the diffusion stages performed alternatively. The confusion and diffusion effect is solely contributed by the permutation and the diffusion stage, respectively. As a result, more overall rounds than necessary are required to achieve a certain level of security. In this paper, we suggest to introduce certain diffusion effect in the confusion stage by simple sequential add-and-shift operations. The purpose is to reduce the workload of the time-consuming diffusion part so that fewer overall rounds and hence a shorter encryption time is needed. Simulation results show that at a similar performance level, the proposed cryptosystem needs less than one-third the encryption time of an existing cryptosystem. The effective acceleration of the encryption speed is thus achieved.

## 1. Introduction

Nowadays, communication networks such as mobile networks and the Internet are well developed. However, they are public networks and are not suitable for the direct transmission of confidential messages. To make use of the communication networks already developed and to keep the secrecy simultaneously, cryptographic techniques need to be applied. Traditional symmetric ciphers such as Data Encryption Standard (DES) are designed with good confusion and diffusion properties [1]. These two properties can also be found in chaotic systems which are usually ergodic and are sensitive to system parameters and initial conditions. In recent years, a number of chaos-based cryptographic schemes have been proposed. Some of them are based on one-dimensional chaotic maps and are applied to data sequence or document encryption [2-4]. For image encryption, two-dimensional (2D) or higher-dimensional chaotic maps are naturally employed as the image can be considered as a 2D array of pixels [5-12].

In [5], Fridrich suggested that a chaos-based image encryption scheme should compose of two processes: chaotic confusion and pixel diffusion. The former permutes the pixels of a plain image with a 2D chaotic map while the latter alternates the value (gray-level) of each pixel in a sequential manner. This architecture formed the basis of a number of chaos-based image ciphers proposed subsequently. For example, Chen and his research group employed a three-dimensional (3D) cat map [10] and a 3D baker map [11] in the confusion stage. Guan *et al* used a 2D cat map for pixel position permutation and the discretized Chen's chaotic system for pixel value masking [7].

In [8], Lian *et al* pointed out that there exist some weak keys for ciphers employing the cat and the baker maps. Moreover, the key space of these two maps is not as large as that of the standard map. Therefore they suggested using a standard map for confusion while keeping the logistic map for pixel value diffusion. To achieve a satisfactory level of security, Lian *et al* recommended to perform four overall rounds of confusion and diffusion [8]. In each confusion stage, 4 permutation rounds should be performed. These lead to a total of 16 permutation rounds and 4 diffusion rounds. Although measures such as pre-computation of permutation mode and sine table were suggested to reduce the computational complexity, the relatively slow diffusion process still limits the performance of this cryptosystem.

To accelerate the encryption speed of Lian *et al*'s cryptosystem and other ciphers based on the iterative confusion-diffusion processes, we propose to introduce certain diffusion effect in the confusion process so that this effect is not solely contributed by the slow diffusion process. Simulation results show that the number of overall rounds and hence the number of time-consuming diffusion processes is reduced without sacrificing the security level. The overall encryption time is shortened although the time required in the confusion stage is increased slightly.

The paper is organized as follows. In the next section, the architecture of cryptosystems based on iterative confusion-diffusion processes is introduced with Lian *et al*'s scheme [8] as an example. The design concept of our approach is described in Section 3. Simulation results and performance analyses are reported in Section 4. In the last section, a conclusion is drawn.

## 2. Architecture of Chaos-based Image Cryptosystems

A typical architecture of the chaos-based image cryptosystems is shown in Fig. 1.

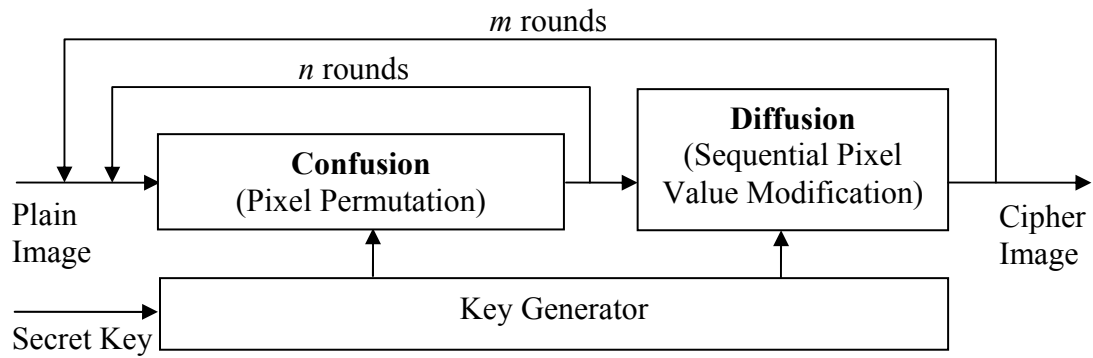

Fig. 1. Typical architecture of the chaos-based image cryptosystems.

There are two iterative stages in the chaos-based image cryptosystem. The confusion stage permutes the pixels in the image, without changing its value. In the diffusion stage, the pixel values are modified sequentially so that a tiny change in one pixel is spread out to many pixels, hopefully the whole image. To decorrelate the relationship between adjacent pixels, there are *n* permutation rounds in the confusion stage with $n \geq 1$. The whole confusion-diffusion round repeats for a number of times to achieve a satisfactory level of security. The parameters of the chaotic maps governing the permutation and the diffusion should better be different in different rounds. This is achieved by a round key generator with a seed secret key as input.

In Lian *et al*'s cryptosystem [8], the confusion process is realized solely by permuting pixel positions without pixel value mixing. It employs an invertible discretized 2D standard map with the introduction of random scan couple ($r_x$, $r_y$) for corner-pixel confusion, as given by Eq. (1).

$$x_{k+1} = (x_k + y_k + r_x + r_y) \bmod N,$$
$$y_{k+1} = \left( y_k + r_y + K_C \sin \frac{2\pi x_{k+1}}{N} \right) \bmod N, \qquad (1)$$

where ($x_k$, $y_k$) and ($x_{k+1}$, $y_{k+1}$) is the original and the permuted pixel position of an $N \times N$ image, respectively. The standard map parameter $K_C$ is a positive integer.

In the diffusion stage, each pixel of the 2D permuted image is scanned sequentially, usually start from the upper left corner. The diffusion effect in this stage is achieved by Eq. (2).

$$\begin{cases} c_{-1} = K_d \\ c_i = v_i \oplus q[f(c_{i-1}), L] \end{cases} \qquad (2)$$

where $v_i$ is the value of the $i^{th}$ pixel of the permuted image, $c_{i-1}$ and $c_i$ is the value of the $(i-1)^{th}$ and the $i^{th}$ pixel of the diffused image, respectively. The seed of the diffusion function is $c_{-1}$ which is obtained from the diffusion key $K_d$. The nonlinear function $f(.)$ is the logistic map given by Eq. (3).

$$f(c_{i-1}) = 4c_{i-1}(1 - c_{i-1}) \qquad (3)$$

The quantization function $q(.)$ takes the $L$ bits just after the decimal point, as defined by Eq. (4).

$$q(X, L) = 2^L \cdot X \qquad (4)$$

where $X = 0.b_1 b_2 b_3 \dots b_L \dots$ is the binary representation of $X$ and $b_i$ is either 0 or 1.

The new pixel value is obtained by exclusive-OR (XOR) the current pixel value $v_i$ of the permuted image with an *L*-bit sequence obtained from the logistic map taking the previous diffused pixel value $c_{i-1}$ as input. As the previous diffused pixel will affect the current one, a tiny change in the plain image is reflected in more than one pixel in the cipher image and so the diffusion effect is introduced in this stage. To generate the distinct confusion and diffusion keys used in different rounds, a key generator composed of skew tent maps is suggested in [8].

## 3. The Proposed Scheme

The proposed scheme is a modification of the one suggested by Lian *et al*.[8]. In their cryptosystem, an explicit diffusion function based on a logistic map is used to spread out the influence of a single plain image pixel over many cipher image elements. Although the diffusion function is executed at a fairly high rate, it is still the highest cost, in terms computational time, of the whole cryptosystem. This is because multiplications of real numbers are required in this stage. Table 1 lists the time required in different parts of Lian *et al*'s cryptosystem. It shows that the time for completing a single diffusion round is more than four times longer than that for a permutation. The test is performed on a personal computer (PC) with 3GHz Pentium D processor, 512 MB memory and 80GB harddisk capacity.

Table 1
Time required in different parts of Lian *et al*'s cryptosystem for encrypting the Lena image sized 512 x 512 with 256 gray levels.

| Task | Key Generation | Single Permutation | Single Diffusion |
|---|---|---|---|
| Time Required (*ms*) | 0.06 | 12.76 | 52.8 |

The encryption speed can be accelerated substantially if fewer diffusion rounds are required. However, the diffusion effect is downgraded if we simply reduce the number of diffusion rounds and keep other parts unchanged. A better way is to introduce certain diffusion effect in the permutation stage as well. The architecture of the proposed cryptosystem is shown in Fig. 2. In the confusion stage, both the permutation on pixel position and the change of pixel value are carried out at the same time while the diffusion process remains unchanged. As a result, the pixel value mixing effect of the whole cryptosystem is contributed by two levels of diffusing operations: the modified confusion process and the original diffusion function. As the diffusion effect is not solely contributed by the diffusion function, the same level of security is achieved in fewer cipher rounds. The encryption speed is thus accelerated.

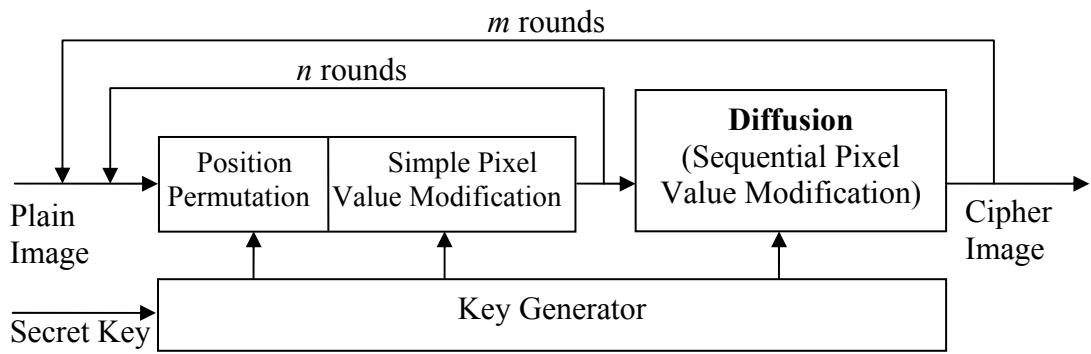

Fig. 2. Architecture of the proposed chaos-based image cryptosystem.

In our modified confusion stage, the new position of a pixel is calculated according to Eq. (1). Before performing the pixel relocation, diffusion effect is injected by adding the current pixel value of the plain image to the previous permuted pixel and then performs a cyclic shift. Other simple logical operations such as XOR can be used instead of the addition operation. The shift operation can also be performed before addition. However, simulation results found that the "add and then shift" combination leads to the best performance and so it becomes the choice in our cryptosystem. The new pixel value is then given by Eq. (5).

$$v_i = Cyc[(p_i + v_{i-1}) \bmod L, \quad LSB_3(v_{i-1})] \tag{5}$$

where $p_i$ is the current pixel value of the plain image, $L$ is the total number of gray levels of the image, $v_{i-1}$ is the value of the $(i-1)^{th}$ pixel after permutation, $Cyc[s, q]$ is the $q$-bit right cyclic shift on the binary sequence $s$, $LSB_3(s)$ refers to the value of the least three significant bits of $s$, $v_i$ is the resultant pixel value in the permuted image. The seed $v_{-1} \in [0, L-1]$ is a sub-key obtained from the key generator.

Similar to the effect of using higher-dimensional chaotic maps for image encryption [10, 11], our modification makes the histogram of confused image uniform in a few rounds. Take a 512×512 white square image as an example of homogeneous image.

Our cryptosystem gives a noisy cipher image and a uniform histogram in only three confusion rounds, as shown in Figs. 3(a) and (b), respectively. By the property of pixel-dependent pixel value mixing, the value of every single pixel is diffused over the entire image. It is regarded as the first level diffusion of the proposed cryptosystem.

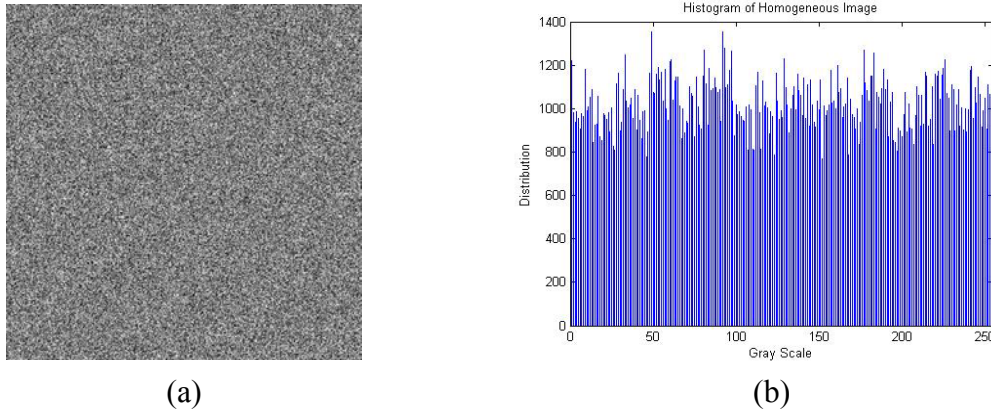

(a)                                (b)

Fig. 3: A white homogenous image processed by 3 rounds of our modified confusion stage. (a) the resultant "noisy" cipher image; (b) the corresponding histogram.

As the pixel value mixing depends on the value of the previously processed pixel, the operation sequence cannot be changed. This may lead to a problem in the reversed confusion process required in decryption. A solution is to make the first decipher round perform the reverse position permutation only. Then both reverse position permutation and pixel value change are performed from the second decipher round. In this manner, an additional decipher round is required for the reverse of pixel value modification. It composes of the simple add-and-shift operation and adds only little cost to the overall decryption procedures.

As mentioned in Section 2, an explicit chaotic diffusion function based on the logistic map is employed in Lian *et al*'s cryptosystem. It is preserved in ours as the second level diffusion for achieving a higher diffusion rate. Similar to Lian *et al*'s cryptosystem for higher security requirements [8,12], the plain image is firstly processed by the modified confusion operation for *n* rounds, followed by a diffusion round. Then the whole process is repeated for *m* rounds.

## 4. Experimental results

Simulation results and performance analyses of the proposed image encryption scheme are provided in this section. They indicate that, at a similar performance level, the proposed cryptosystem leads to a higher encryption speed than Lian *et al*'s.

Firstly, the performance of the confusion methods in Lian *et al*'s and our schemes are compared. The plain image Lena of size $512 \times 512$ and 256 gray levels is employed. It is shown in Fig. 4(a) while its histogram is given in Fig. 4(b). After three rounds of the confusion process, the intermediate cipher image obtained by Lian *et al*'s confusion method is shown in Fig. 4(c). As only pixel permutation is performed by their confusion, the corresponding statistical information depicted in Fig. 4(d) is exactly the same as that of the plain image. Figure 4(e) is the cipher image obtained by the proposed confusion scheme. The corresponding histogram found in Fig. 4(f) indicates a promising degree of pixel value mixing in only 3 confusion rounds.

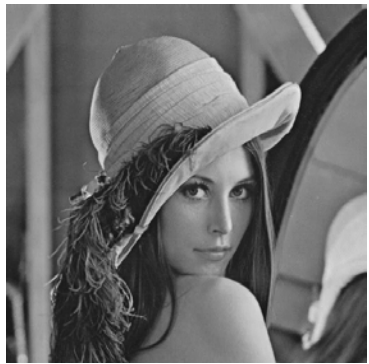
(a)

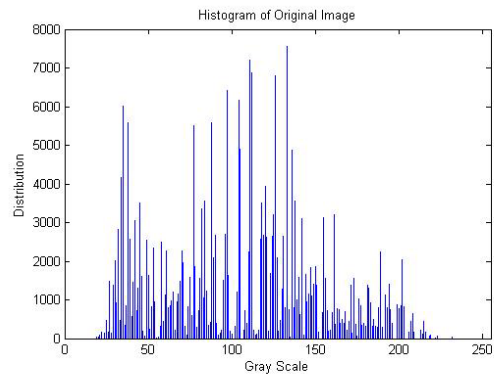
(b)

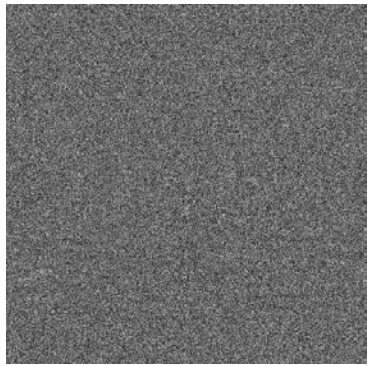
(c)

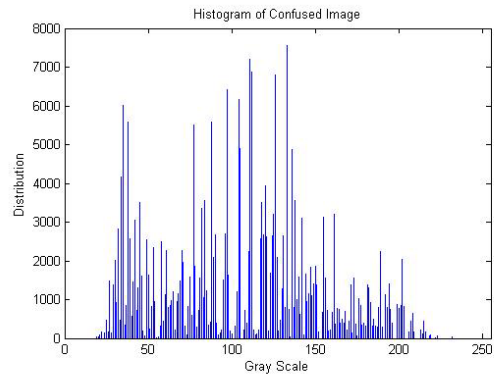
(d)

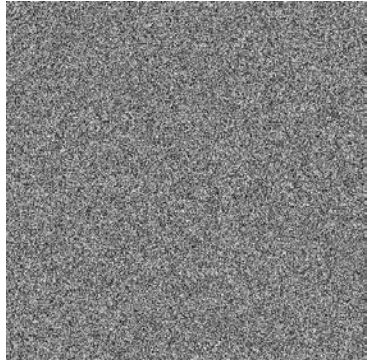 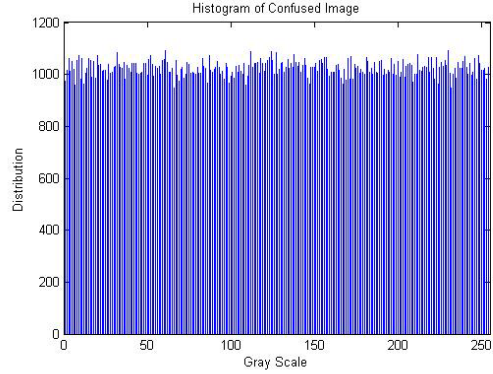

(e) (f)

Fig. 4. (a) Plain Lena image; (b) Histogram of the plain image; (c) Intermediate cipher image using Lian *et al*'s confusion; (d) Histogram of the intermediate cipher image shown in (c); (e) Intermediate cipher image using the proposed confusion; (f) Histogram of the intermediate cipher image given in (e).

In addition to the above simulation, an experiment dedicated to the confusion process has been carried out. The results shown in Fig. 5 demonstrate that our proposed confusion method is sensitive to two plain images different by only one bit. Figure 5(a) is the original Camerman image of size 256×256 in 256 gray levels. Figures 5(b) and (c) are the cipher images obtained after 3 rounds of the proposed confusion process, whose corresponding plain images have only a 1 bit difference at the lower right corner. The two cipher images have 99.36% of pixel different with each other. The difference image between the two cipher images can be found in Fig. 5(d). Such results are benefited from the pixel value modification introduced in the confusion process. During the pixel position permutation, the gray level values of all pixels of the image are changed pixel-by-pixel depending on their neighborhood. This acts as the first level diffusion in our scheme. As a result, the one-bit difference in the plain image diffuses substantially over many pixels in the cipher image.

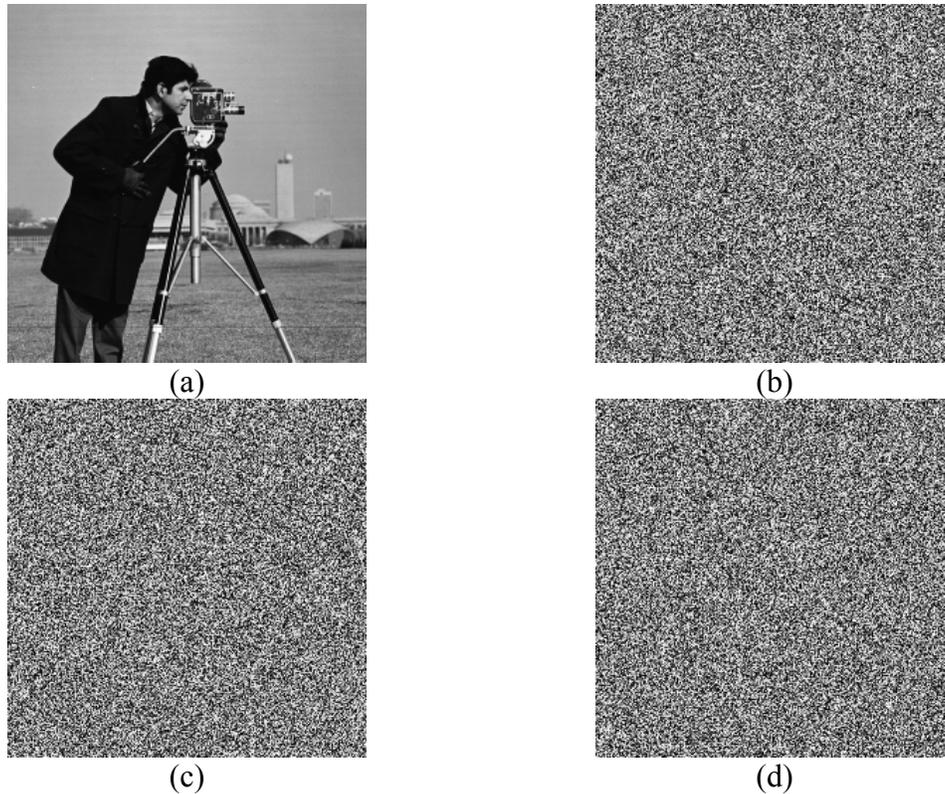

Fig. 5. (a) Plain Cameraman image; (b) and (c) cipher images whose corresponding plain images have one-bit difference only; (d) difference between cipher images shown in (b) and (c).

Indeed, the diffusion effect introduced in the confusion process supplements that contributed by the explicit diffusion function. Therefore our cryptosystem achieves a similar performance in fewer cipher rounds than Lian *et al*'s. This is supported by the simulation results of the proposed and Lian *et al*'s schemes at different combinations of confusion ($n$) and overall ($m$) rounds using the 512×512 Lena image in 256 gray scales. Table 2 lists the encryption and decryption time required in both schemes. The computer configuration used in this test is again a 3 GHz Pentium D processor with 512 MB memory and 80G harddisk capacity. In addition to the encryption time, two performance indices, namely, number of pixels change rate (NPCR) and unified average changing intensity (UACI) as adopted in [10,11] are also listed in Table 2. Their trend at different overall rounds with $n$ fixed to 4 is plotted in Figs. 6(a) and (b), respectively. The graphs show that both performance indices rise rapidly in our proposed scheme, which indicate good confusion and diffusion effect.

The simulation results listed in Table 2 show that to achieve a similar performance of Lian *et al*'s recommended cryptosystem ($m=n=4$) [8], the proposed scheme only requires one overall round with three permutation rounds in each confusion stage, i.e., $m=1$ and $n=3$. The corresponding encryption time is 108.73 milliseconds (*ms*) which is just a quarter of Lian *et al*'s (468.89 *ms*). To achieve a higher performance such as NPCR>0.996 and UACI>0.334, Lian *et al*'s requirement is $m=6$ and $n=3$ while the proposed scheme only needs $m=2$ and $n=2$. Our encryption time (189.03 *ms*) is less than one-third of Lian *et al*'s (626.46 *ms*). The substantial acceleration in encryption speed is due to the reduction of the number of overall rounds $m$. Fewer time-consuming diffusion operations are needed and so the encryption time is shortened. The additional computation complexity of the simple add-and-shift operation in the modified confusion stage is insignificant. It leads to an extra encryption time of only 1.5 *ms* per permutation, as given by the difference between the two encryption time data in the first row ($m=1$ and $n=2$) of Table 2. The decryption time for both the proposed and Lian *et al* 's schemes are longer than the corresponding encryption time. However, the increase is not substantial, only 4.8% to 9.1 % in our scheme and 0.2% to 4.7% in Lian *et al*'s scheme, as computed from the decryption time data listed in Table 2.

Table 2. Encryption time and performance indices NPCR and UACI of the proposed and Lian *et al*'s schemes, for some selected values of *m* and *n*.

| (*m*, *n*) | Encryption Time (*ms*) | | Decryption Time (*ms*) | | NPCR | | UACI | |
|---|---|---|---|---|---|---|---|---|
| | Proposed | Lian *et al* | Proposed | Lian *et al* | Proposed | Lian *et al* | Proposed | Lian *et al* |
| (1,2) | 94.48 | 91.58 | 103.08 | 95.55 | 0.668423 | 0.000179 | 0.202745 | 0.000040 |
| (1,3) | **108.73** | 104.23 | 118.29 | 109.11 | **0.994007** | 0.000252 | **0.327571** | 0.000061 |
| (1,4) | 122.86 | 117.91 | 132.25 | 121.57 | 0.995941 | 0.000423 | 0.334354 | 0.000093 |
| (2,2) | **189.03** | 183.65 | 202.92 | 187.84 | **0.996117** | 0.011070 | **0.334731** | 0.002750 |
| (2,3) | 217.87 | 209.12 | 233.64 | 212.92 | 0.996147 | 0.017376 | 0.334249 | 0.004581 |
| (2,4) | 247.27 | 234.33 | 263.87 | 239.22 | 0.996227 | 0.022896 | 0.334245 | 0.006024 |
| (3,2) | 282.73 | 274.55 | 303.87 | 278.96 | 0.995956 | 0.438816 | 0.334844 | 0.119449 |
| (3,3) | 330.04 | 313.54 | 347.15 | 317.89 | 0.996136 | 0.566250 | 0.335066 | 0.155397 |
| (3,4) | 369.90 | 353.27 | 389.54 | 356.08 | 0.996269 | 0.639618 | 0.335371 | 0.175316 |
| (4,2) | 377.11 | 370.39 | 403.16 | 371.70 | 0.996117 | 0.980228 | 0.335290 | 0.298463 |
| (4,3) | 434.63 | 418.87 | 461.97 | 421.84 | 0.996159 | 0.987320 | 0.334295 | 0.306969 |
| (4,4) | 496.06 | **468.89** | 521.09 | 472.94 | 0.996243 | **0.989960** | 0.334756 | **0.310990** |
| (4,5) | 550.39 | 521.72 | 577.34 | 525.92 | 0.995884 | 0.947609 | 0.334564 | 0.275717 |
| (5,3) | 543.52 | 522.97 | 576.31 | 525.44 | 0.996113 | 0.995907 | 0.335271 | 0.331217 |
| (5,4) | 615.79 | 587.42 | 647.56 | 592.50 | 0.996166 | 0.995750 | 0.335041 | 0.330743 |
| (5,5) | 688.34 | 652.52 | 721.65 | 653.54 | 0.996037 | 0.995304 | 0.333939 | 0.326393 |
| (6,2) | 567.27 | 549.30 | 604.65 | 556.26 | 0.995861 | 0.995991 | 0.334693 | 0.333462 |
| (6,3) | 653.00 | **626.46** | 689.35 | 631.78 | 0.996216 | **0.996197** | 0.334953 | **0.334627** |

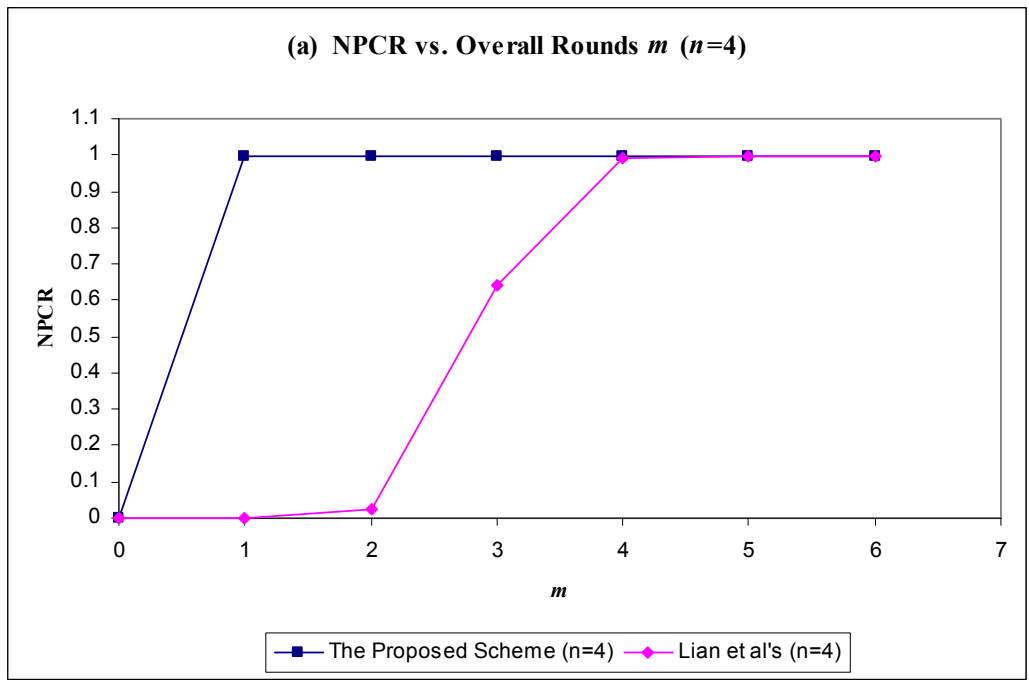

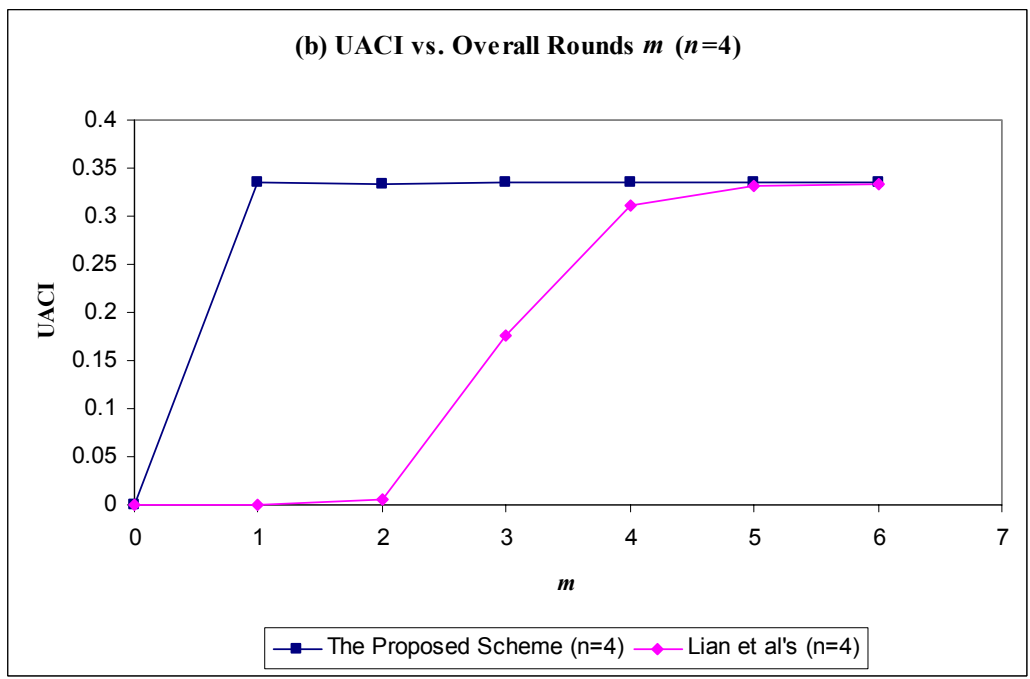

Fig. 6. Performance of the proposed and Lian *et al*'s cryptosystems in terms of (a) number of pixels change rate (NPCR); and (b) unified average changing intensity (UACI) at different overall rounds (*m*) with 4 permutation rounds in each confusion stage (*n* = 4).

In general, adjacent pixels of most plain images are highly correlated. However, one of the requirements of an effective image cryptosystem is the generation of a cipher image with sufficiently low correlation of adjacent pixels. To analyze the effectiveness of our cryptosystem in this aspect, the correlations between two adjacent pixels in horizontal, vertical and diagonal directions are calculated. In the experiment, four images, namely, a 256 gray scale plain Peppers image of size $512 \times 512$, the cipher images obtained using the proposed scheme ($m = 2$ and $n = 2$) and Lian *et al*'s scheme ($m = 6$ and $n = 3$), and a randomly generated test image are employed. The correlation coefficients are calculated by the formula stated in [10,11] and are listed in Table 3. The data for the two cipher images are in the same order of magnitude as those for the random image. This implies that both cryptosystems can effectively decorrelate adjacent pixels in the plain image. As an example, the correlation distribution of two horizontally adjacent pixels of the plain image and the cipher image obtained using the proposed scheme is shown in Figs. 7(a) and 7(b), respectively.

Table 3. Correlation coefficients of adjacent pixels of different images.

|  | Plain Peppers image | Cipher image by the proposed scheme ($m=n=2$) | Cipher image by Lian *et al*'s scheme ($m=6, n=3$) | Random image |
|---|---|---|---|---|
| Horizontal | 0.982208 | 0.002637 | 0.002453 | 0.001562 |
| Vertical | 0.977046 | 0.009177 | 0.004864 | 0.005922 |
| Diagonal | 0.978133 | 0.003429 | 0.007525 | 0.004006 |

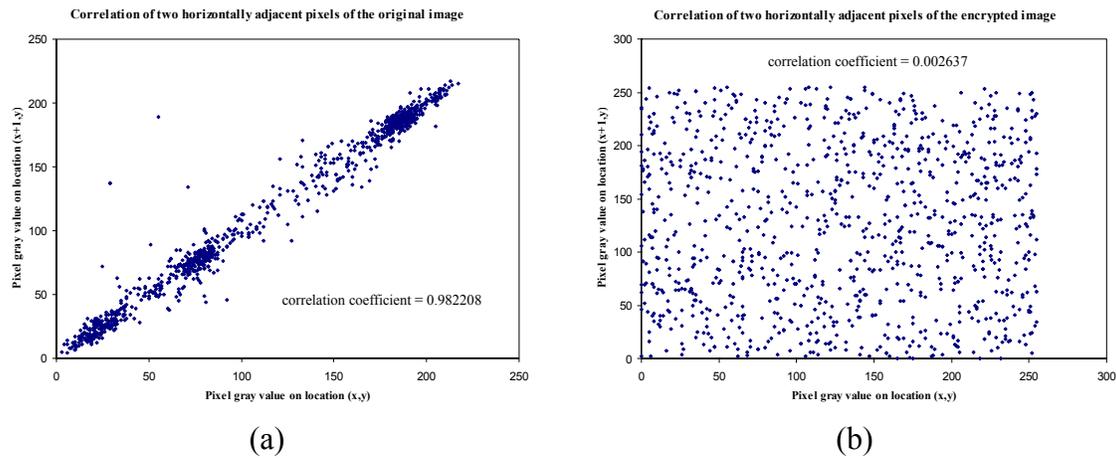

(a)                                    (b)

Fig. 7. Correlation analysis of two horizontally adjacent pixels in (a) the plain Peppers image; (b) the cipher image obtained using the proposed scheme.

## 5. Conclusions

The typical structure of chaos-based image encryption schemes has been studied. It is found that the diffusion part is the most time-consuming one as multiplications of real numbers are required in the logistic map. To reduce the overall computational complexity for accelerating the encryption speed, we suggest introducing certain diffusion effect in the permutation stage by simple sequential add-and-shift operations. Simulation results show that at a similar performance level, the proposed cryptosystem needs less than one-third encryption time as that of Lian *et al*'s. The effective sharing of the workload in the time-consuming diffusion part is achieved.

**Acknowledgement**

The work described in this paper was fully supported by a grant from CityU (Project no. 7001927).